\documentclass{aastex631}
\makeatletter

\newcommand{\Rmnum}[1]{\expandafter\@slowromancap\romannumeral #1@}
\makeatother

\shorttitle{Instantaneous heating events}
\shortauthors{Yang et al.}

\graphicspath{{./}{figures/}}

\begin{document}

\title{Atmospheric Heating Events Associated with Fine-scale Flux Emergence in Ephemeral Regions}

\author{Hanlin Yang}
\affiliation{State Key Laboratory of Solar Activity and Space Weather, National Astronomical Observatories, Chinese Academy of Science}
\affiliation{University of Chinese Academy of Sciences, 100049, Beijing, People’s Republic of China}

\author{Jingxiu Wang}
\affiliation{State Key Laboratory of Solar Activity and Space Weather, National Astronomical Observatories, Chinese Academy of Science}
\affiliation{University of Chinese Academy of Sciences, 100049, Beijing, People’s Republic of China}

\author{Chunlan Jin}
\affiliation{State Key Laboratory of Solar Activity and Space Weather, National Astronomical Observatories, Chinese Academy of Science}

\begin{abstract}

Coronal heating has puzzled solar physicists for decades. The question of why the Sun’s upper atmosphere is significantly hotter than its lower atmosphere remains a key mystery. It is commonly believed that the source of coronal heating comes from the Sun's magnetic field, and more complex magnetic dynamics is more efficient in heating. In an earlier work we studied the secondary (or finer-sacle) flux emergence identified in five ephemeral regions (ERs), selected during the last solar minimum \citep{2024ApJ...967...59Y}. Here we further explore the atmospheric response to the secondary flux emergences (SFEs) that were identified in the first paper. We further reveal that approximately 80 percent of the 172 identified SFEs are associated with atmospheric heating events. The heating is most likely associated with magnetic reconnection involved in the SFE. Overall, a solar quiet region is heated by several hundred thousand degrees, during flux emergence of an ER. 
%In addition, we observed a few microflares and ejections with clear atmospheric heating, which were associated with tiny magnetic features of flux density lower than the detection limit.
\end{abstract}


\section{Introduction} \label{sec:intro}

The solar corona, a tenuous outer atmosphere of the Sun, presents one of the astrophysics’ most enduring mysteries—how it maintains temperatures exceeding a million Kelvin while the underlying photosphere is orders of magnitude cooler. This phenomenon, termed as the coronal heating problem, has spurred extensive observational and theoretical investigations. \citet{2006SoPh..234...41K} articulated a systematic framework to tackle this challenge, emphasizing the roles of wave dissipation, and magnetic reconnection as potential contributors. Later, \citet{2012RSPTA.370.3217P} highlighted the critical importance of small-scale magnetic phenomena, such as the magnetic carpet, in reconciling energy budgets and localized heating processes. Historically, observations from space-based observatories like Yohkoh, TRACE, and SOHO have revealed dynamic structures in the corona, such as loops and oscillations, offering insights into magnetohydrodynamic (MHD) processes \citep{2010LRSP....7....5R}. Furthermore, advanced numerical simulations have underscored the efficacy of nanoflares—small-scale reconnection events initially theorized by \citet{1988ApJ...330..474P}—as a potential universal heating mechanism \citep{1994ApJ...422..381C}. The interplay of waves and turbulent dissipation is another hypothesis, with observational support for Alfvén waves being supplied from the chromosphere to the corona \citep{2020SSRv..216..140V}. Despite substantial progress, challenges persist in differentiating the dominant heating mechanisms, especially across varying solar conditions.

Investigations into this longstanding problem have remained ongoing. Parker’s nanoflare hypothesis \citep{1988ApJ...330..474P} proposed that small, discrete reconnection events driven by tangled magnetic field lines could release energy at scales sufficient to maintain coronal heating. This concept was later refined through numerical simulations, which highlighted the necessity of efficient energy conversion and deposition mechanisms \citep{1994ApJ...422..381C, 2002ApJ...572L.113G}. Concurrently, wave-based heating mechanisms, particularly those involving Alfvén waves, gained attraction as models demonstrated their ability to transport energy from the photosphere to the corona \citep{2005SSRv..120...67O, 2011Natur.475..477M}. 

In parallel, flux emergence has gained attention as a key contributor to energy deposition in the solar atmosphere. Using 3D MHD simulations, \citet{2014PASJ...66...39L} demonstrated that the twist and expansion properties of emerging magnetic loops significantly influence both coronal heating and solar wind acceleration. \citet{2014A&A...564A..12C} further illustrated that magnetic braiding driven by flux emergence leads to current formation and coronal loop heating. Quantitative analyses by \citet{2015PASJ...67...18W} and \citet{2015RSPTA.37340263L} found that the Poynting flux and reconnection rates in flux emergence regions can meet the energetic requirements of coronal heating.
These theoretical findings have been increasingly supported by observations. \citet{2018A&A...612A..28L} reported enhanced chromospheric heating co-spatial with emerging flux regions, based on high-resolution spectropolarimetric data. More recently, \citet{2021ApJ...911...41G} presented direct evidence that internetwork magnetic fields can ascend from the photosphere into the transition region, contributing to localized heating with fluxes up to $5.3 \times 10^{18}$ Mx and plasma flows of 10 km s$^{-1}$.

Despite decades of research, the coronal heating problem remains unresolved, partly due to the observational difficulty of disentangling these processes in a highly dynamic and multi-scale plasma environment.
%Recent advances indicate that magnetic flux emergence from the photosphere profoundly influences coronal heating. 
High-resolution imaging from the Solar Dynamics Observatory (SDO) and numerical models support the hypothesis that magnetic flux tubes, interacting through braiding and reconnection, can generate sufficient energy to heat the corona \citep{2013A&A...555A.123B}. The dynamics of magnetic flux emergence, coupled with interactions with the pre-existing coronal magnetic field, have emerged as a critical area of investigation. As convective motions in the solar interior drive flux emergence, they induce localized perturbations and reconnection events that contribute significantly to coronal energy deposition \citep{2014LRSP...11....3C}.

Ephemeral regions, characterized by short-lived, small-scale magnetic flux emergence, are particularly important in this context. Observations from Hinode and SDO have linked such regions with localized heating events, including upward-propelled plasma jets and magnetic reconnection \citep{2011LRSP....8....6S}. \citet{2019A&A...626A..98L} further demonstrated that flux emergence correlates with episodic heating, showing that the interplay between upward-propelled jets and reconnection can deposit energy sufficient to balance radiative and conductive losses. These findings underpin our investigation into the connection between magnetic emergence and episodic coronal heating.
%Ephemeral regions, characterized by short-lived, small-scale magnetic flux emergence, are particularly important in this context. Observations from Hinode and SDO have linked such regions with localized heating events, including upward-propelled plasma jets and magnetic reconnection \citep{2011LRSP....8....6S, 2002A&ARv..10..313P}. Recent studies using high-resolution magnetograms and multi-wavelength imaging have provided compelling evidence that photospheric vortex motions—twisting and stressing magnetic field lines—trigger energy release in the overlying corona \citep{2012Natur.486..505W}. \citet{2019A&A...626A..98L} further demonstrated that flux emergence correlates with episodic heating, showing that the interplay between upward-propelled jets and reconnection can deposit energy sufficient to balance radiative and conductive losses. These findings underpin our investigation into the connection between magnetic emergence and episodic coronal heating.

In our earlier work \citep{2024ApJ...967...59Y}, we demonstrated that ephemeral regions are far from simple bipolar magnetic structures. Instead, they involve multiple episodes of secondary flux emergences (SFEs) over their lifetimes. Across the five ERs, we identify 172 SFEs. Notably, most occur during the early phase of the ERs, aligning with the period of rapid total magnetic flux increase of ERs. The discovery of the complex magnetic evolution process naturally raises the question: do these intricate flux emergences induce significant atmospheric heating and, if so, in what ways and to what extent?

In this study, we examined five ephemeral regions observed during the solar minima of Cycles 23 and 24 \citep{2024ApJ...967...59Y}. Our observational results reveal that as much as approximately 80\% of SFEs are associated with heating events that elevate the atmospheric temperature to over 3 MK. Moreover, for the ephemeral regions as a whole, magnetic flux emergence leads to a substantial increase in the mean atmospheric temperature of the studied quiet region, rising from a background level of approximately 2 MK to 2.6 MK. Interestingly, the evolution of the average atmospheric temperature is not a steady increase or decrease. Instead, it exhibits a pulsed pattern of rises and falls, reflecting the episodic nature of energy release during the flux emergences.

We organize the remainder of this paper as follows: In Section \ref{sec:ob}, we detail our observational setup, including the instrumentation, data acquisition, and analysis methods employed. In Section \ref{sec:results}, we present our key findings. Finally, in Section \ref{sec:conclusion}, we summarize and discuss these results.

\section{Observation} \label{sec:ob}

The five ephemeral regions (ERs) appear between 2010.11.20 12:00:00 UT and 2010.11.25 12:00:00 UT, located in very quiet region within latitudes $ \pm 37^\circ $ and longitudes $ \pm 25^\circ $ on the solar disk. We utilize magnetograms from Helioseismic and Magnetic Imager (HMI) onboard SDO to analyze the structure and evolution of magnetic field. Atmospheric Imaging Assembly (AIA) observations are employed to investigate the corresponding radiative properties in the overlying atmosphere. We adopt methods described in \citet{2001ApJ...555..448H}, \citet{2004SoPh..219...39L} and \citet{2011ApJ...731...37J} to evaluate magnetic field noise. This approach involves performing a Gaussian fit to the frequency distribution of magnetic flux densities across all pixels during the quiescent phase of the ephemeral region. The standard deviation is determined to be 7 Gauss. Thus, we set the magnetic field detection threshold at $3\sigma$, corresponding to 21 Gauss. This threshold ensures robust detection of the magnetic flux elements, minimizing spurious contributions from noise. 

To analyze the thermal properties of the ephemeral regions using emission measure (EM), we utilize the Differential Emission Measure (DEM) method by \citet{2015ApJ...807..143C}. DEM quantifies the distribution of plasma along different temperatures, which is critical for assessing the localized heating. The DEM is defined as:

\begin{equation}
DEM(T) = n_e^2 \frac{dh}{dT},
\end{equation}
where $ n_e $ is the electron density, $ h $ is the column height along the line of sight, and $ T $ is the plasma temperature. This formulation reflects the plasma’s contribution to the observed emission as a function of temperature.
The DEM inversion employs six AIA channels: $ 94 $$ \text{\AA} $, $ 131 $$ \text{\AA} $, $ 171 $$ \text{\AA} $, $ 193 $$ \text{\AA} $, $ 211 $$ \text{\AA} $, and $ 335 $$ \text{\AA} $. These channels are chosen because their passbands correspond to specific temperature ranges, providing sensitivity to plasma spanning $ 10^5 $$ \text{K} $ to $ 10^7 $$ \text{K} $. For instance, the $ 171 $$ \text{\AA} $ channel is most sensitive to cooler plasmas around 0.6$\text{MK}$, while $ 131 $$ \text{\AA} $ and $ 94 $$ \text{\AA} $ effectively capture hotter components around $ 10 $$ \text{MK} $. $ 304 $$ \text{\AA} $ channel is excluded from the DEM inversion for $304$$ \text{\AA} $ channel primarily observing He II lines, which are optically thick. This optical thickness complicates the assumptions of line-of-sight integration used in DEM inversions, leading to potential inaccuracies.

We employ linear programming techniques to ensure positive-definite DEM solutions, as detailed in the Cheung et al. (2015) method. To compute the emission measure (EM), we divide the temperature range into 20 logarithmically spaced bins between $ 10^5 $$ \text{K} $ and $ 10^7 $$ \text{K} $. This binning approach balances resolution with stability in the DEM inversion. The EM is defined as:

\begin{equation}
EM = \int_{T_{\text{min}}}^{T_{\text{max}}} DEM(T) dT,
\end{equation}
where $ T_{\text{min}} $ and $ T_{\text{max}} $ correspond to the bounds of each temperature bin. By summing contributions over these bins, we derive the total EM for each region, allowing for detailed mapping of coronal plasma densities across temperature ranges.

The temperature ($T_{EM}$) of the coronal plasma is derived directly from the DEM solution:
\begin{equation}
T_{EM} = \frac{\int_{T_{\text{min}}}^{T_{\text{max}}} T \cdot DEM(T) dT}{\int_{T_{\text{min}}}^{T_{\text{max}}} DEM(T) dT},
\end{equation}
where: $T$ is the plasma temperature at a given layer. The DEM quantifies the amount of plasma along the line of sight as a function of temperature, and the temperature is obtained as a weighted average using DEM distribution.
Such temperatures allow us to pinpoint regions of enhanced heating and associate them with small-scale magnetic activities.

For each ER, we classify regions containing the majority of magnetic flux and undergoing gradual, visibly distinct topological changes as main polarities. Small bipoles occurring within or near these main polarities are defined as SFEs. These small positive polarities and negative polarities emerge adjacent to each other, separate with each other and eventually merge with main polarities, becoming indistinguishable. We visually identify and track the SFEs to determine whether or not they are associated with atmospheric heating events. In our previous study \citep{2024ApJ...967...59Y}, we classified SFEs into four types based on their emergence locations and morphological characteristics: the regular arch type, kink type, twist type, and bubble type.
For the regular arch type, SFEs typically emerge between the main polarities and often develop into major flux elements within the emerging region (ER). The kink type exhibits an axis opposite to that of the main polarities, with one polarity undergoing cancellation with the ambient magnetic field. The twist type shares the same axial orientation as the main polarities, yet one polarity still cancels with the surrounding flux. The bubble type lacks a preferred orientation, and both polarities eventually dissipate. This classification provides a clearer picture for understanding the evolution of SFEs.

\section{Results} \label{sec:results}
\begin{figure}[h]
\centering
\includegraphics[width=1.0\textwidth]{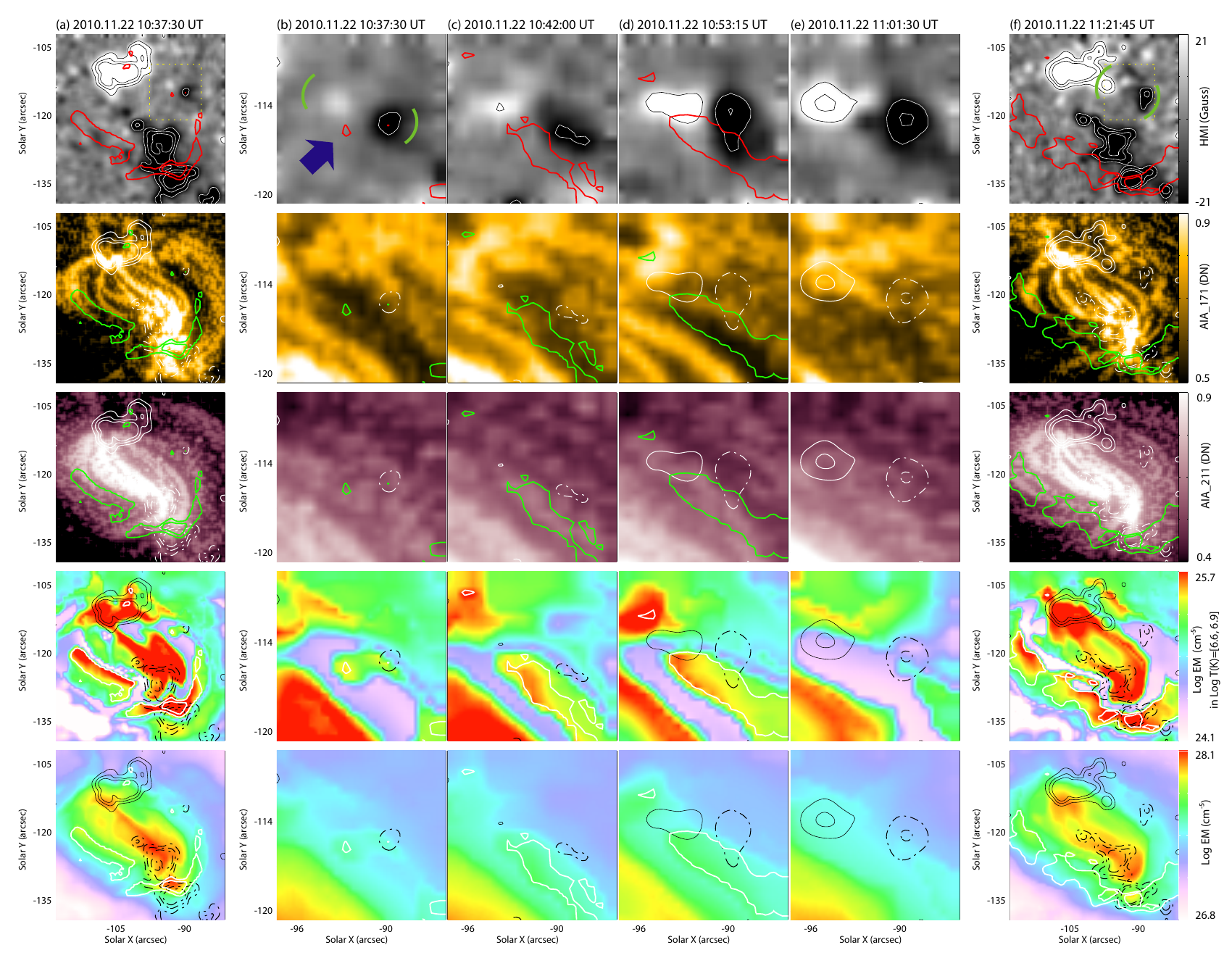}
\caption{Heating events triggered by SFEs. From top to bottom: line-of-sight magnetic field, AIA-171, AIA-211, EM in log T (K)=[6.6, 6.9] and total EM. Regions exceeding 3 MK are outlined in red on the magnetic field map, green on the AIA-171 and AIA-211 maps, and white on the EM map. The positive and negative polarity contours are represented by black and white solid lines in the magnetogram, white solid and dashed lines in the AIA-171 and AIA-211 images, and black solid and dashed lines in the EM map. Panels (b)-(e) refer to areas within yellow box of panel (a). Green brackets in the magnetic field map mark the location of the SFE, while the blue arrow indicates the heating region. The AIA-171 $\text{\AA}$ and AIA-211 $\text{\AA}$ panels in this figure adopts the plotting method described by \citet{2014SoPh..289.2945M}, with the parameter set to $\text{h=0.92}$, to highlight the structure and shape of the coronal loop as clearly as possible. \label{fig:emerge1}} 
\end{figure}

We systematically examine each SFE and its associated radiative response to better understand the dynamic coupling between magnetic flux emergence and atmospheric heating. Figure \ref{fig:emerge1} illustrates an atmospheric heating event triggered by an SFE in ER 1. From left to right, the panels are arranged in chronological order. From top to bottom, they display line-of-sight magnetic field, AIA-171, AIA-211, emission measure (EM) in log T (K)=[6.6, 6.9] and total EM. Here, we include AIA-211 images for comparison. Since AIA-211 has a characteristic temperature of approximately 2 MK \citep{2012SoPh..275...17L}, which is close to the average temperature of the ER upper atmosphere, we investigate whether regions of atmospheric heating coincide with brightening in the AIA-211 images. To highlight heating activity clearly, regions with temperatures exceeding 3 MK are outlined in red contours on the magnetic field map, green on the AIA-171 and AIA-211 maps, and white on the EM map.
Panels (b)–(e) show the region outlined by the yellow box in panel (a). At 10:37:30 UT, a small bipole, marked by green brackets, begins to emerge. The negative polarity exceeds the detection threshold, while the positive polarity remains below it. However, as the bipole emerges, parts of the overlying atmosphere reach 3 MK, indicated by the blue arrow in the magnetogram. After 4 min 30 sec, the high-temperature region expands, forming a strip-like structure above the bipole. Meanwhile, the central area of the positive polarity surpasses 21 Gauss. 11 min 15 sec later, the emergence region further expands, and atmospheric heating continues. However, after 8 min 15 sec, the 3 MK region disappears, while the bipole fully emerges and separates. At 11:21:45 UT, the small positive polarity merges into the main positive polarity.
Given the strong temporal and spatial correlation between this heating event and the SFE, we attribute the heating to the SFE. However, we find that in the log EM maps within the temperature range of log $T = [6.6, 6.9]$, the enhanced regions align well with the locations of the heating event. This indicates that 3 MK heating events are accompanied by significant EM enhancement in the 4--8 MK temperature range. Additionally, in the regions of peak temperature, the radiative intensity in the AIA-171 (0.6 MK) and AIA-211 (6.3 MK) images is not the strongest. For example, in panel (c), the highest-temperature region shows weaker AIA-211 emission than the area to its east, which remains below 3 MK. This contrast is even more evident in the AIA-171 images. Conversely, the regions with strong radiation in AIA-171 and AIA-211 correspond to areas of enhanced total EM but not to the high temperatures. This indicates that heating events cannot be solely analyzed through increased radiation.

\begin{figure}[h]
\centering 
\includegraphics[width=1.0\textwidth]{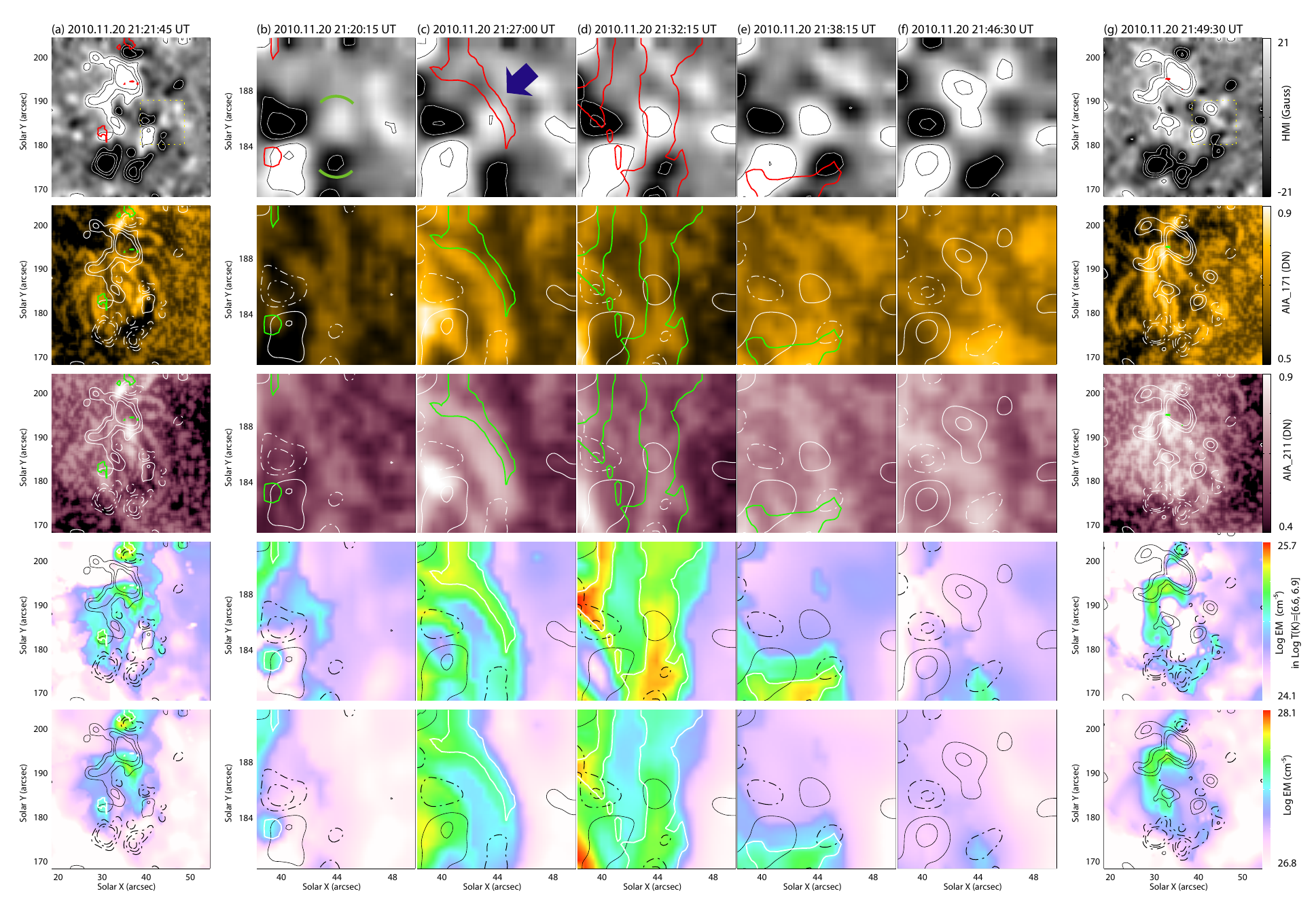}
\caption{Heating events triggered by SFEs. From top to bottom: line-of-sight magnetic field, AIA-171, AIA-211, EM in log T (K)=[6.6, 6.9] and total EM. Regions exceeding 3 MK are outlined in red on the magnetic field map, green on the AIA-171 and AIA-211 maps, and white on the EM map. The positive and negative polarity contours are represented by black and white solid lines in the magnetogram, white solid and dashed lines in the AIA-171 and AIA-211 images, and black solid and dashed lines in the EM map. Panels (b)-(e) refer to areas within yellow box of panel (a). Green brackets in the magnetic field map mark the location of the SFE, while the blue arrow indicates the heating region.  \label{fig:emerge2}}
\end{figure}

Similar to Figure \ref{fig:emerge1}, Figure \ref{fig:emerge2} also presents an atmospheric heating event associated with an SFE in ER 2. At 21:20:15 UT, a small bipole marked by green brackets begins to emerge. The negative polarity is already visible with contour lines, while the positive polarity remains below the detection threshold. After 6 min and 45 sec, the positive polarity surpasses the threshold. As it emerges, the overlying atmosphere heats up to 3 MK.
The heated region continues to expand, and by 21:32:15 UT, it fully covers the area above the SFE. Shortly after, the hot region starts contracting. By 21:38:15 UT, only the atmosphere above the negative polarity remains above 3 MK. Over the next 8 min and 15 sec, the heating event gradually subsides as the bipole separates. Finally, at 21:49:30 UT, the negative polarity of the SFE, marked by green brackets, merges into the main negative polarity.
The strong spatiotemporal correlation between the heating event and the evolution of the SFE suggests that the heating is driven by the SFE. In this event, we note that the region of high-temperature activity still shows significant enhancement in the log EM map within the range of log $T = [6.6, 6.9]$. This case also reveals partial structural correspondence in the AIA-171 images. This resembles the bidirectional brightenings observed by \citet{2024A&A...692A.119C} along the arch-filament system (AFS). However, the plasma in our event reaches temperatures up to 3 MK, while the temperatures derived by \citet{2024A&A...692A.119C} are only around $10^{5.5}$ K. This difference may indicate that, in Figure~\ref{fig:emerge2}, the plasma rises to higher altitudes and interacts with environments at greater heights, leading to enhanced energy release and heating.

\begin{figure}[h]
\centering
\includegraphics[width=1.0\textwidth]{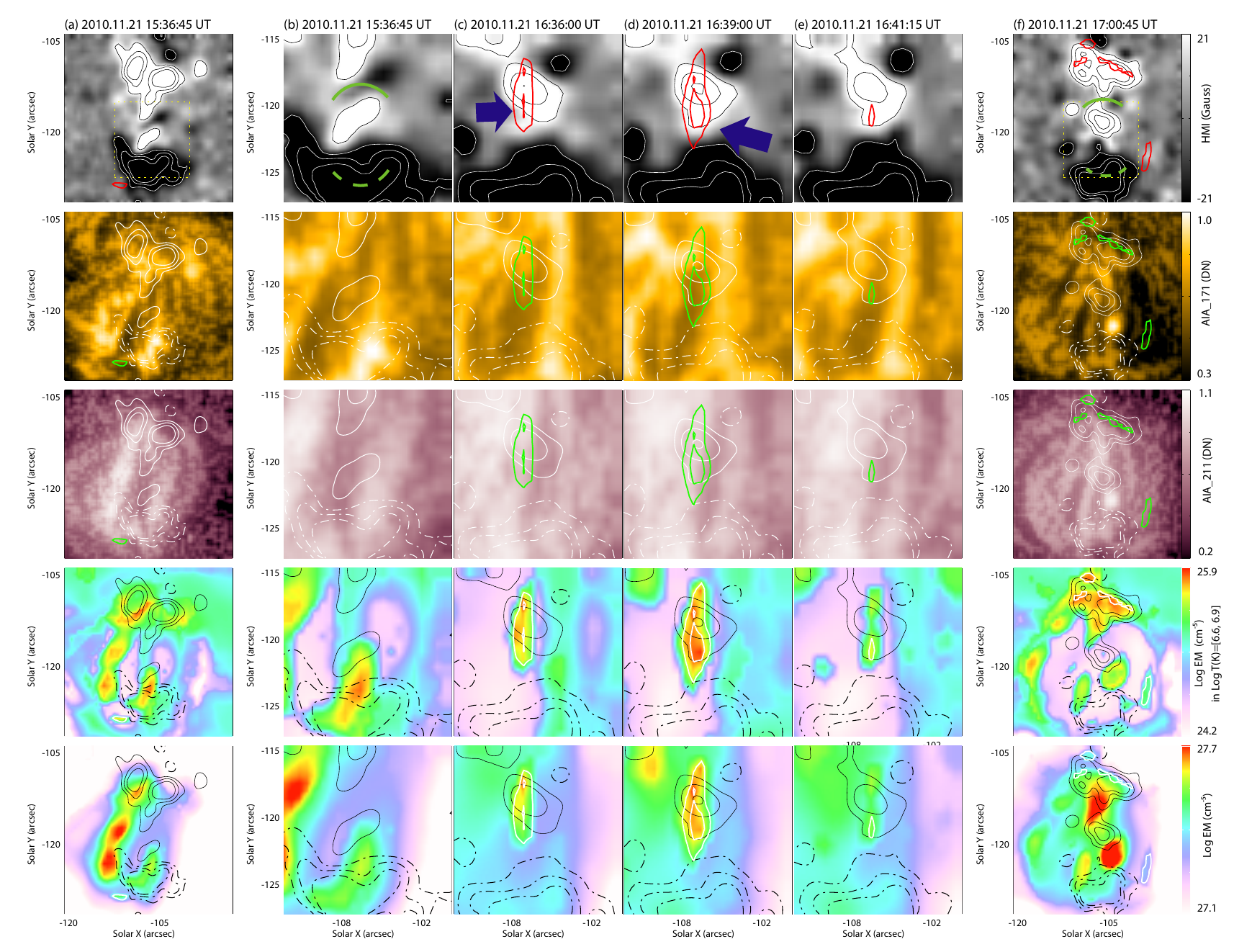}
\caption{Heating events triggered by SFEs. From top to bottom: line-of-sight magnetic field, AIA-171, AIA-211, EM in log T (K)=[6.6, 6.9] and total EM. Regions exceeding 3 MK are outlined in red on the magnetic field map, green on the AIA-171 and AIA-211 maps, and white on the EM map. The positive and negative polarity contours are represented by black and white solid lines in the magnetogram, white solid and dashed lines in the AIA-171 and AIA-211 images, and black solid and dashed lines in the EM map. Panels (b)-(e) refer to areas within yellow box of panel (a). Green brackets in the magnetic field map mark the location of the SFE, while the blue arrow indicates the heating region.  \label{fig:emerge3}}
\end{figure}

In some cases, not both polarities of an SFE could be simultaneously observable, as one polarity is embedded within the main polarity and remains indistinguishable at the onset. Figure \ref{fig:emerge3} presents a heating event triggered by an SFE where only one polarity is visible. At 15:36:45 UT, the positive polarity emerges, surpassing the 21 Gauss contour, marked by green semi-brackets. However, the negative polarity remains unresolved, because it is too close to the main negative polarity, or the spatial resolution is insufficient. We infer that it is likely hidden within the main negative polarity, indicated by the green dashed brackets. The small positive polarity continues to emerge, gradually increasing in magnetic flux density.
An hour later, at 16:36:00 UT, a high-temperature region exceeding 3 MK appears directly above the small positive polarity, with small areas surpassing 4 MK, as indicated by the blue arrow. Three minutes later, the positive polarity emerges further, while the 3 MK and 4 MK regions expand. The central temperature peaks at 4.5 MK. After 2 minutes and 15 seconds, the 3 MK region starts contracting. By 17:00:45 UT,  the heating event within the yellow box ceases entirely, and the positive polarity fully merges into the main positive polarity. Although the negative polarity of the SFE remains unresolved, we deduce that it is likely embedded within the main negative polarity, as indicated by the dashed brackets. Additionally, we observe that, in both AIA-171 and AIA-211, the regions with the strongest radiation are not always the hottest. For example, in panel (b), the second and third rows show intense radiation in AIA-171 and AIA-211, but no region exceeds 3 MK. Similarly, the EM maps in panels (a) and (f) reveal that the regions with the highest EM do not correspond to the hottest regions. DEM method better highlights coronal high-temperature regions, tracking the occurrence and intensity of heating events.

\begin{figure}[h]
\centering
\includegraphics[width=1.0\textwidth]{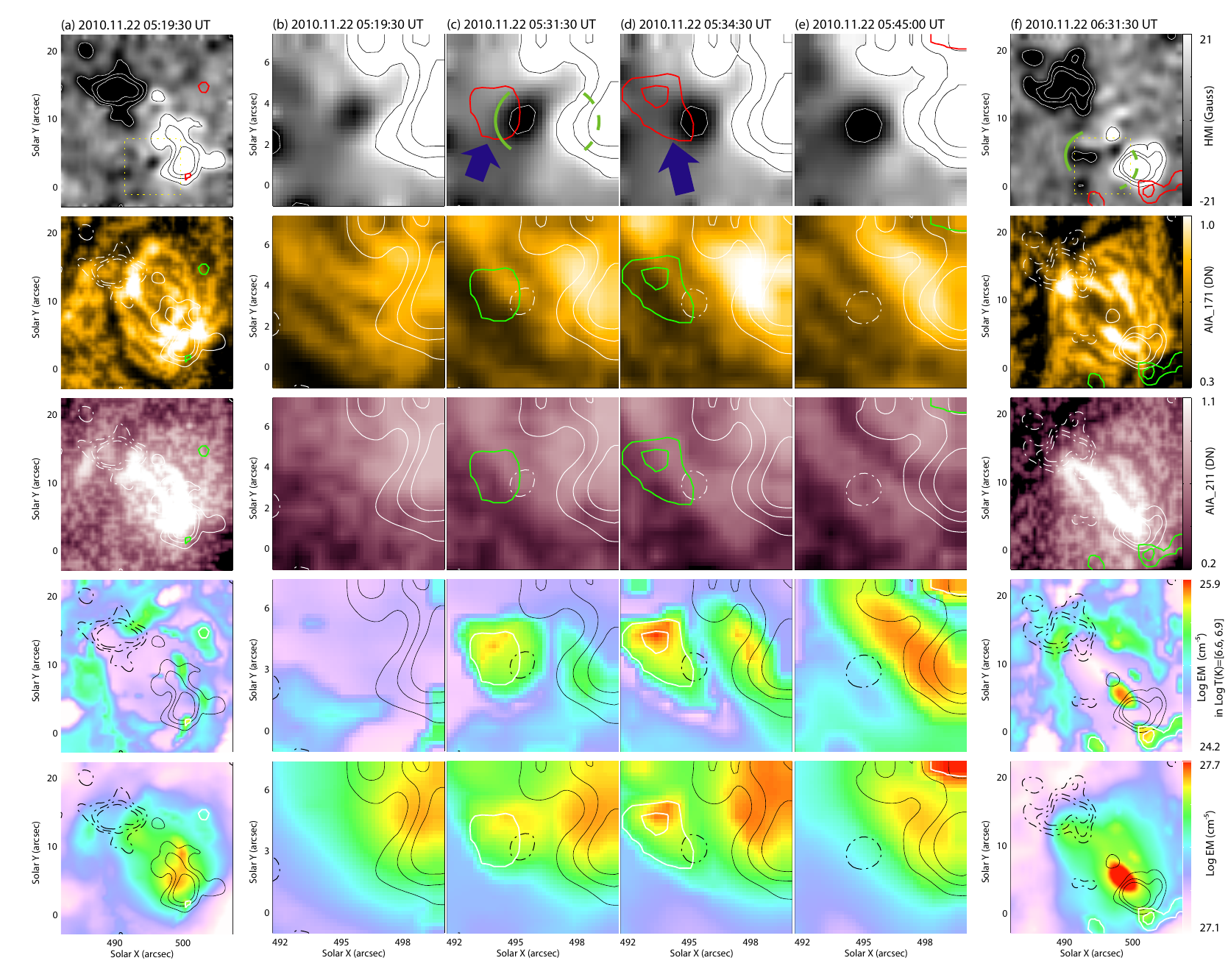}
\caption{Heating events triggered by SFE. From top to bottom: line-of-sight magnetic field, AIA-171, AIA-211, EM in log T (K)=[6.6, 6.9] and total EM. Regions exceeding 3 MK are outlined in red on the magnetic field map, green on the AIA-171 and AIA-211 maps, and white on the EM map. The positive and negative polarity contours are represented by black and white solid lines in the magnetogram, white solid and dashed lines in the AIA-171 and AIA-211 images, and black solid and dashed lines in the EM map. Panels (b)-(e) refer to areas within yellow box of panel (a). Green brackets in the magnetic field map mark the location of the SFE, while the blue arrow indicates the heating region.  \label{fig:emerge4}} 
\end{figure}

Another typical example is shown in Figure \ref{fig:emerge4}. Within the yellow box in panel (a), a small negative polarity below the detection threshold emerges. At 05:31:30 UT, it surpasses the 21 Gauss threshold, marked by green semi-brackets. However, no corresponding positive polarity is observed. We infer that it is likely embedded within the main positive polarity, indicated by the green dashed lines.
A high-temperature region exceeding 3 MK appears in the atmosphere above, as shown by the blue arrow. Three minutes later, the heated area expands further, reaching 4 MK at its center, with a peak temperature of 4.3 MK. However, the heating event is short-lived and ceases entirely by 05:45:00 UT. Afterward, the small negative polarity moves away from the main positive polarity and shifts toward the main negative polarity, as illustrated in panel (f). This case is particularly interesting because the negative polarity of the SFE originates within the main positive polarity, partially disrupting its magnetic feature.

\begin{figure}[h]
\centering
\includegraphics[width=1.0\textwidth]{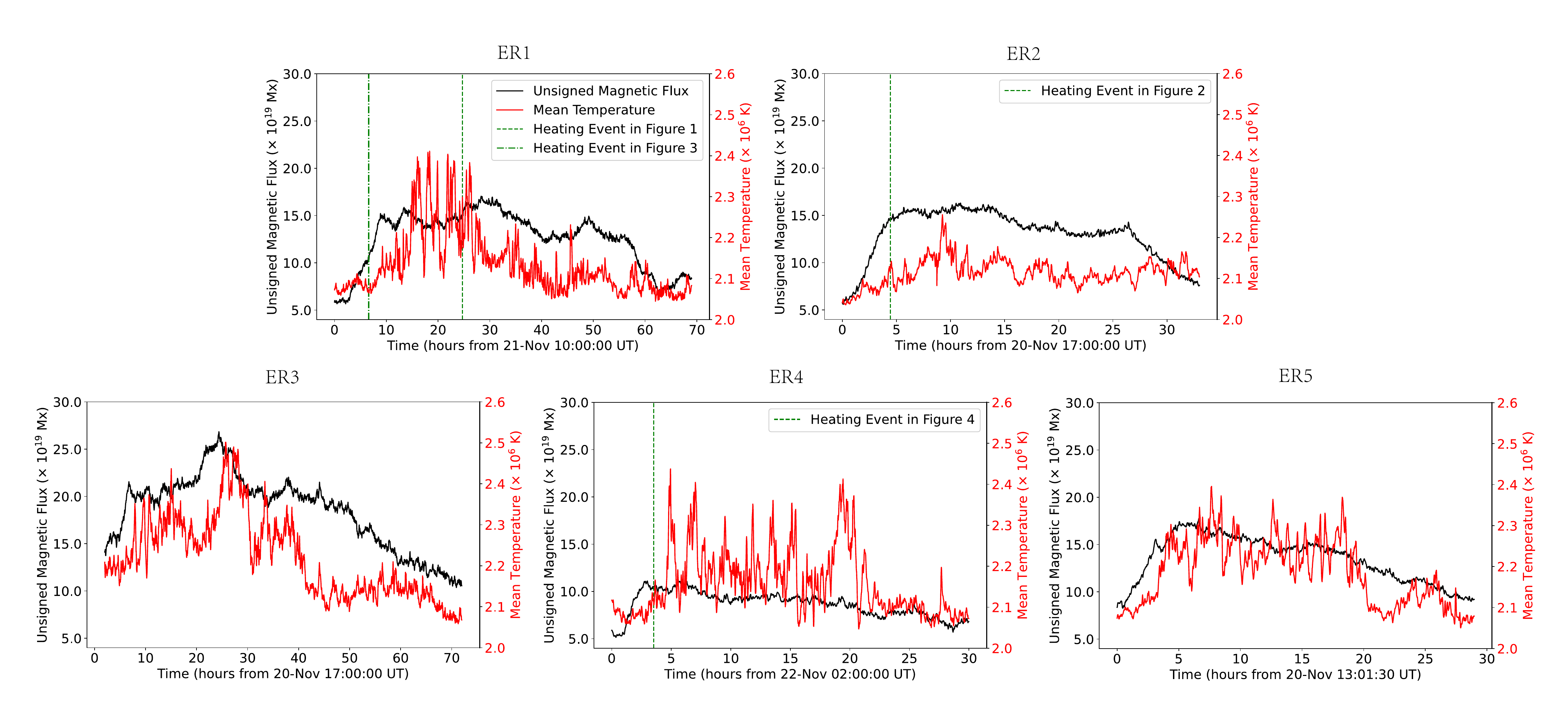}
\caption{Temperature variation synchronized with magnetic flux in five ERs. The black line represents unsigned magnetic flux, and the red line indicates the average temperature. The green dashed lines mark the time snapshots of the heating events shown in Figures \ref{fig:emerge1}-\ref{fig:emerge4}. The left vertical axis corresponds to magnetic flux, while the right vertical axis shows temperature. \label{fig:tem}} 
\end{figure}

After examining each SFE, we find that 134 out of 172 identifiable SFEs in the five ERs trigger atmospheric heating events during their evolution, accounting for 77.9 \%.
Naturally, we investigate whether the overall flux emergence in an ER influences the average temperature of the region. To determine this, we plot the magnetic flux and average temperature for each of the five ERs as line graphs, shown in Figure \ref{fig:tem}. For each ER, we define the study area based on its evolution stage when both the magnetic field area and the EM area reach their maximum. At this time, a rectangular region with a field of view 1.5 times the length and width of the ER area is selected. Within this region, we measure the total magnetic flux and the average temperature.
The results indicate that flux emergence in ERs generally leads to an increase in the average temperature during their evolution. For example, in ER1, the peak of flux emergence raises the average temperature by approximately 0.4 MK compared to its minimum. In contrast, in ER2, the temperature increases by only about 0.2 MK during flux emergence.
Interestingly, the magnitude of flux emergence does not always correspond to the extent of temperature rise, as seen by comparing ER2 and ER4. The extent of temperature rise seems primarily linked to the structural complexity of the magnetic field. However, overall, we observe that as the magnetic flux reaches its maximum, the average temperature in the ER also peaks. As the flux declines, the temperature gradually returns to a lower level. We also find that the average temperature is highly sensitive to how the  field of view is defined. A slightly larger selection will cause the temperature curve to flatten rapidly, losing variations and approaching 2 MK. 
It is noteworthy that, compared to the gradual and low-level variations in magnetic flux, the average temperature exhibits significant and abrupt fluctuations over short periods. The temperature changes are not gradual but occur in a pulsated manner. This observation aligns with the description by \citet{2015RSPTA.37340256K}, where coronal heating is described as transient and characterized by impulsive energy release.

\section{Conclusion and discussion}\label{sec:conclusion}

Through detailed observations of five ERs at the solar disk center from 2010.11.20 12:00:00 UT to 2010.11.25 12:00:00 UT, we identify numerous heating events associated with SFEs. A total of 77.90\% of SFEs are found to trigger atmospheric heating events. The emergence and evolution of ERs also influence the regional average temperature, causing increases ranging from 0.2 MK to 0.4 MK. In the absence of external environmental effects, the average temperature rises with flux emergence and decreases as flux dissipates. However, no evidence suggests a direct correlation between the magnitude of flux emergence and the extent of temperature increase. %Instead, temperature rise appears to be more closely linked to the complexity of the magnetic field. 
%Future studies with additional cases could help uncover potential undiscovered patterns.

In the study by \citet{2022ApJ...929..103T}, the authors investigated small-scale bright dots in the 174~\AA\ passband within a emerging flux region. They reported that these brightenings were associated with mixed-polarity magnetic flux emergence and cancellation in the photosphere, and originated from plasma at temperatures of approximately 0.8--1.2~MK---indicative of the lower corona or transition region.
In contrast, our observations reveal heating events at coronal heights that significantly exceed the typical average coronal temperature ($\sim$3~MK). These high-temperature events are similarly driven by SFEs within or in the vicinity of ERs, but differ in that they are not always accompanied by brightening in the AIA 171~\AA\ or 211~\AA\ channels; in some cases, emission in these passbands is even depressed relative to the surrounding background. This difference may suggest that the two phenomena occur at different atmospheric heights, with our observations corresponding to higher coronal layers. This motivates the use of direct DEM inversion to identify such heating events more reliably.

Our observations suggest that flux emergence-induced atmospheric heating events fall into three categories.
The first type, similar to the event in Figure \ref{fig:emerge1} and \ref{fig:emerge2}, seems to occur when newly emerging flux tubes inject plasma and energy into an overlying emerging arch, heating the entire magnetic structure. These events often lead to the heating of large-scale loops spanning the main polarities of the ER. They affect a broader region, persist for a longer duration, and are particularly notable because small-scale SFEs can trigger extensive atmospheric heating.
The second type, as shown in Figures \ref{fig:emerge3}, appears to result from magnetic reconnection between newly emerged loops and preexisting coronal fields at the loop apex, leading to localized energy release. These events typically cover a small area, last for a little more than ten minutes, and occur between main positive and negative polarities. After a brief, pulse-like heating phase, they dissipate. This type is most commonly associated with SFEs during the early stages of ER evolution.
The third type, like event in Figure \ref{fig:emerge4}, is likely caused by flux cancellation between SFE and main polarities. These events are typically short-lived, confined to a small region, and occur near a single main polarity. According to the classification of SFEs \citep{2024ApJ...967...59Y}, the events presented in Figures~\ref{fig:emerge1}--\ref{fig:emerge3} belong to the regular arch type, whereas the case in Figure~\ref{fig:emerge4} is categorized as the twist type. Future high-resolution observations may provide deeper insights into the underlying physical mechanisms and magnetic structure evolution of these heating processes.

In \citet{2024ApJ...967...59Y}, we reported that SFEs predominantly occur in the early evolutionary phase of ERs. Since SFEs induce localized atmospheric heating, we are also interested in how ER-scale flux emergence affects temperature. To explore this, we compare magnetic flux and average temperature in Figure \ref{fig:tem}. It shows a time lag between flux and temperature peaks in each ER. The magnetic flux always reaches its maximum first, followed by the peak in average temperature. Magnetic flux initially emerges at the photosphere. The newly emerged flux then rises into the corona, where it interacts with the ambient magnetic field. This interaction leads to observable responses in coronal images, from which we infer the associated temperature changes. Naturally, there exists a temporal delay between the magnetic emergence at the photosphere and the coronal response. This delay may also result from the complex restructuring of the overlying magnetic field during flux emergence, including SFEs. Such complexity enhances magnetic energy release through reconnection and its conversion into thermal energy. As a result, temperature responds slightly later than flux growth. During ER evolution, heating caused by SFEs is only one of many heating processes. We also observe various small-scale activities, such as microflares and plasma ejections, triggering strong atmospheric responses. Some magnetic loops even exhibit temperatures above 3 MK immediately upon emergence. Future studies with additional cases could help uncover potential undiscovered patterns.

The emergence of low-level magnetic flux from the solar interior into the atmosphere is (or sometimes) a critical process influencing atmospheric heating. Magnetic reconnection, triggered as the emerging magnetic flux interacts with preexisting coronal fields, facilitates energy dissipation and plasma heating. For example, \citet{2004A&A...426.1047A} conducted numerical experiments demonstrating that reconnection at the interface of newly emerged and ambient magnetic fields generates significant heating in the solar corona. Similarly, \citet{2008A&A...479..567M} showed that the atmospheric response to emerging flux tubes includes the formation of current sheets and dissipation of magnetic energy, leading to localized temperature increases. These studies underscore the pivotal role of low-level magnetic flux emergence in driving energy deposition and heating in the solar atmosphere.

Small-scale activity phenomenon is another significant source of atmospheric heating. These localized events, often linked to small-scale magnetic reconnection, deposit energy into the lower solar atmosphere, creating multi-thermal plasma structures. \citet{2017ApJ...839...22H} provided evidence from 3D radiative MHD simulations that reconnection between emerging bipolar magnetic fields produces localized heating, extending from the photosphere to the upper chromosphere. 
%Similarly, \citet{2014LRSP...11....3C} highlighted the importance of reconnection-driven energy release during flux emergence, explaining the observed temperature enhancements in transition region and chromospheric layers. 
Furthermore, \citet{2017ApJ...836...63T} examined early-stage flux emergence in emerging flux regions, showing that localized reconnection events lead to intense, short-lived brightenings associated with atmospheric heating.

These findings collectively suggest that low-level magnetic flux emergence and small-scale reconnection phenomena play critical roles in heating the solar atmosphere. They offer valuable insights into the dynamic coupling between magnetic fields and plasma in the Sun’s outer layers. Future studies combining high-resolution observations with advanced numerical models will further elucidate these complex processes and their implications for solar atmospheric dynamics.

SDO is the first mission to be launched for NASA's Living With a Star Program. The authors are grateful to the team members who have made great contributions to the SDO mission. This work is supported by the B-type Strategic Priority Program of the Chinese Academy of Sciences (grant No. XDB0560000), the National Key R\&D Program of China (grant No. 2021YFA1600500 ), and the National Natural Science Foundation of China (grant No. 12273061 and 12350004). This work is supported by the Specialized Research Fund for State Key Laboratory of Solar Activity and Space Weather.

\bibliography{sample631}{}

\begin{thebibliography}{}
\expandafter\ifx\csname natexlab\endcsname\relax\def\natexlab#1{#1}\fi
\providecommand{\url}[1]{\href{#1}{#1}}
\providecommand{\dodoi}[1]{doi:~\href{http://doi.org/#1}{\nolinkurl{#1}}}
\providecommand{\doeprint}[1]{\href{http://ascl.net/#1}{\nolinkurl{http://ascl.net/#1}}}
\providecommand{\doarXiv}[1]{\href{https://arxiv.org/abs/#1}{\nolinkurl{https://arxiv.org/abs/#1}}}

\bibitem[{{Archontis} {et~al.}(2004){Archontis}, {Moreno-Insertis},
  {Galsgaard}, {Hood}, \& {O'Shea}}]{2004A&A...426.1047A}
{Archontis}, V., {Moreno-Insertis}, F., {Galsgaard}, K., {Hood}, A., \&
  {O'Shea}, E. 2004, \aap, 426, 1047, \dodoi{10.1051/0004-6361:20035934}

\bibitem[{{Bourdin} {et~al.}(2013){Bourdin}, {Bingert}, \&
  {Peter}}]{2013A&A...555A.123B}
{Bourdin}, P.~A., {Bingert}, S., \& {Peter}, H. 2013, \aap, 555, A123,
  \dodoi{10.1051/0004-6361/201321185}

\bibitem[{{Cargill}(1994)}]{1994ApJ...422..381C}
{Cargill}, P.~J. 1994, \apj, 422, 381, \dodoi{10.1086/173733}

\bibitem[{{Chen} {et~al.}(2014){Chen}, {Peter}, {Bingert}, \&
  {Cheung}}]{2014A&A...564A..12C}
{Chen}, F., {Peter}, H., {Bingert}, S., \& {Cheung}, M.~C.~M. 2014, \aap, 564,
  A12, \dodoi{10.1051/0004-6361/201322859}

\bibitem[{{Chen} {et~al.}(2024){Chen}, {Mandal}, {Peter}, \&
  {Chitta}}]{2024A&A...692A.119C}
{Chen}, Y., {Mandal}, S., {Peter}, H., \& {Chitta}, L.~P. 2024, \aap, 692,
  A119, \dodoi{10.1051/0004-6361/202451069}

\bibitem[{{Cheung} {et~al.}(2015){Cheung}, {Boerner}, {Schrijver}, {Testa},
  {Chen}, {Peter}, \& {Malanushenko}}]{2015ApJ...807..143C}
{Cheung}, M. C.~M., {Boerner}, P., {Schrijver}, C.~J., {et~al.} 2015, \apj,
  807, 143, \dodoi{10.1088/0004-637X/807/2/143}

\bibitem[{{Cheung} \& {Isobe}(2014)}]{2014LRSP...11....3C}
{Cheung}, M. C.~M., \& {Isobe}, H. 2014, Living Reviews in Solar Physics, 11,
  3, \dodoi{10.12942/lrsp-2014-3}

\bibitem[{{Go{\v{s}}i{\'c}} {et~al.}(2021){Go{\v{s}}i{\'c}}, {De Pontieu},
  {Bellot Rubio}, {Sainz Dalda}, \& {Pozuelo}}]{2021ApJ...911...41G}
{Go{\v{s}}i{\'c}}, M., {De Pontieu}, B., {Bellot Rubio}, L.~R., {Sainz Dalda},
  A., \& {Pozuelo}, S.~E. 2021, \apj, 911, 41, \dodoi{10.3847/1538-4357/abe7e0}

\bibitem[{{Gudiksen} \& {Nordlund}(2002)}]{2002ApJ...572L.113G}
{Gudiksen}, B.~V., \& {Nordlund}, {\r{A}}. 2002, \apjl, 572, L113,
  \dodoi{10.1086/341600}

\bibitem[{{Hagenaar}(2001)}]{2001ApJ...555..448H}
{Hagenaar}, H.~J. 2001, \apj, 555, 448, \dodoi{10.1086/321448}

\bibitem[{{Hansteen} {et~al.}(2017){Hansteen}, {Archontis}, {Pereira},
  {Carlsson}, {Rouppe van der Voort}, \& {Leenaarts}}]{2017ApJ...839...22H}
{Hansteen}, V.~H., {Archontis}, V., {Pereira}, T.~M.~D., {et~al.} 2017, \apj,
  839, 22, \dodoi{10.3847/1538-4357/aa6844}

\bibitem[{{Jin} {et~al.}(2011){Jin}, {Wang}, {Song}, \&
  {Zhao}}]{2011ApJ...731...37J}
{Jin}, C.~L., {Wang}, J.~X., {Song}, Q., \& {Zhao}, H. 2011, \apj, 731, 37,
  \dodoi{10.1088/0004-637X/731/1/37}

\bibitem[{{Klimchuk}(2006)}]{2006SoPh..234...41K}
{Klimchuk}, J.~A. 2006, \solphys, 234, 41, \dodoi{10.1007/s11207-006-0055-z}

\bibitem[{{Klimchuk}(2015)}]{2015RSPTA.37340256K}
---. 2015, Philosophical Transactions of the Royal Society of London Series A,
  373, 20140256, \dodoi{10.1098/rsta.2014.0256}

\bibitem[{{Lee} \& {Magara}(2014)}]{2014PASJ...66...39L}
{Lee}, H., \& {Magara}, T. 2014, \pasj, 66, 39, \dodoi{10.1093/pasj/psu020}

\bibitem[{{Leenaarts} {et~al.}(2018){Leenaarts}, {de la Cruz Rodr{\'\i}guez},
  {Danilovic}, {Scharmer}, \& {Carlsson}}]{2018A&A...612A..28L}
{Leenaarts}, J., {de la Cruz Rodr{\'\i}guez}, J., {Danilovic}, S., {Scharmer},
  G., \& {Carlsson}, M. 2018, \aap, 612, A28,
  \dodoi{10.1051/0004-6361/201732027}

\bibitem[{{Lemen} {et~al.}(2012){Lemen}, {Title}, {Akin}, {Boerner}, {Chou},
  {Drake}, {Duncan}, {Edwards}, {Friedlaender}, {Heyman}, {Hurlburt}, {Katz},
  {Kushner}, {Levay}, {Lindgren}, {Mathur}, {McFeaters}, {Mitchell}, {Rehse},
  {Schrijver}, {Springer}, {Stern}, {Tarbell}, {Wuelser}, {Wolfson}, {Yanari},
  {Bookbinder}, {Cheimets}, {Caldwell}, {Deluca}, {Gates}, {Golub}, {Park},
  {Podgorski}, {Bush}, {Scherrer}, {Gummin}, {Smith}, {Auker}, {Jerram},
  {Pool}, {Soufli}, {Windt}, {Beardsley}, {Clapp}, {Lang}, \&
  {Waltham}}]{2012SoPh..275...17L}
{Lemen}, J.~R., {Title}, A.~M., {Akin}, D.~J., {et~al.} 2012, \solphys, 275,
  17, \dodoi{10.1007/s11207-011-9776-8}

\bibitem[{{Li} \& {Peter}(2019)}]{2019A&A...626A..98L}
{Li}, L.~P., \& {Peter}, H. 2019, \aap, 626, A98,
  \dodoi{10.1051/0004-6361/201935165}

\bibitem[{{Liu} {et~al.}(2004){Liu}, {Zhao}, \&
  {Hoeksema}}]{2004SoPh..219...39L}
{Liu}, Y., {Zhao}, X., \& {Hoeksema}, J.~T. 2004, \solphys, 219, 39,
  \dodoi{10.1023/B:SOLA.0000021822.07430.d6}

\bibitem[{{Longcope} \& {Tarr}(2015)}]{2015RSPTA.37340263L}
{Longcope}, D.~W., \& {Tarr}, L.~A. 2015, Philosophical Transactions of the
  Royal Society of London Series A, 373, 20140263,
  \dodoi{10.1098/rsta.2014.0263}

\bibitem[{{McIntosh} {et~al.}(2011){McIntosh}, {de Pontieu}, {Carlsson},
  {Hansteen}, {Boerner}, \& {Goossens}}]{2011Natur.475..477M}
{McIntosh}, S.~W., {de Pontieu}, B., {Carlsson}, M., {et~al.} 2011, \nat, 475,
  477, \dodoi{10.1038/nature10235}

\bibitem[{{Morgan} \& {Druckm{\"u}ller}(2014)}]{2014SoPh..289.2945M}
{Morgan}, H., \& {Druckm{\"u}ller}, M. 2014, \solphys, 289, 2945,
  \dodoi{10.1007/s11207-014-0523-9}

\bibitem[{{Murray} \& {Hood}(2008)}]{2008A&A...479..567M}
{Murray}, M.~J., \& {Hood}, A.~W. 2008, \aap, 479, 567,
  \dodoi{10.1051/0004-6361:20078852}

\bibitem[{{Ofman}(2005)}]{2005SSRv..120...67O}
{Ofman}, L. 2005, \ssr, 120, 67, \dodoi{10.1007/s11214-005-5098-1}

\bibitem[{{Parker}(1988)}]{1988ApJ...330..474P}
{Parker}, E.~N. 1988, \apj, 330, 474, \dodoi{10.1086/166485}

\bibitem[{{Parnell} \& {De Moortel}(2012)}]{2012RSPTA.370.3217P}
{Parnell}, C.~E., \& {De Moortel}, I. 2012, Philosophical Transactions of the
  Royal Society of London Series A, 370, 3217, \dodoi{10.1098/rsta.2012.0113}

\bibitem[{{Reale}(2010)}]{2010LRSP....7....5R}
{Reale}, F. 2010, Living Reviews in Solar Physics, 7, 5,
  \dodoi{10.12942/lrsp-2010-5}

\bibitem[{{Shibata} \& {Magara}(2011)}]{2011LRSP....8....6S}
{Shibata}, K., \& {Magara}, T. 2011, Living Reviews in Solar Physics, 8, 6,
  \dodoi{10.12942/lrsp-2011-6}

\bibitem[{{Tiwari} {et~al.}(2022){Tiwari}, {Hansteen}, {De Pontieu}, {Panesar},
  \& {Berghmans}}]{2022ApJ...929..103T}
{Tiwari}, S.~K., {Hansteen}, V.~H., {De Pontieu}, B., {Panesar}, N.~K., \&
  {Berghmans}, D. 2022, \apj, 929, 103, \dodoi{10.3847/1538-4357/ac5d46}

\bibitem[{{Toriumi} {et~al.}(2017){Toriumi}, {Katsukawa}, \&
  {Cheung}}]{2017ApJ...836...63T}
{Toriumi}, S., {Katsukawa}, Y., \& {Cheung}, M. C.~M. 2017, \apj, 836, 63,
  \dodoi{10.3847/1538-4357/836/1/63}

\bibitem[{{Van Doorsselaere} {et~al.}(2020){Van Doorsselaere}, {Srivastava},
  {Antolin}, {Magyar}, {Vasheghani Farahani}, {Tian}, {Kolotkov}, {Ofman},
  {Guo}, {Arregui}, {De Moortel}, \& {Pascoe}}]{2020SSRv..216..140V}
{Van Doorsselaere}, T., {Srivastava}, A.~K., {Antolin}, P., {et~al.} 2020,
  \ssr, 216, 140, \dodoi{10.1007/s11214-020-00770-y}

\bibitem[{{Welsch}(2015)}]{2015PASJ...67...18W}
{Welsch}, B.~T. 2015, \pasj, 67, 18, \dodoi{10.1093/pasj/psu151}

\bibitem[{{Yang} {et~al.}(2024){Yang}, {Jin}, {Wang}, \&
  {Wang}}]{2024ApJ...967...59Y}
{Yang}, H., {Jin}, C., {Wang}, Z., \& {Wang}, J. 2024, \apj, 967, 59,
  \dodoi{10.3847/1538-4357/ad3947}

\end{thebibliography}
\bibliographystyle{aasjournal}

\end{document}